\pdfoutput=1

\documentclass[twocolumn,10pt]{article}

\usepackage[margin=1in]{geometry}
\usepackage{amsmath,amssymb}
\usepackage{graphicx}
\usepackage{subcaption}
\usepackage{listings}
\usepackage{hyperref}
\usepackage{url}
\usepackage{booktabs}
\usepackage{natbib}

\hypersetup{
    colorlinks=true,
    linkcolor=blue,
    citecolor=blue,
    urlcolor=blue
}

\begin{document}

\title{From Funding to Findings (FIND): An Open Database of NSF Awards and Research Outputs}

\author{
    Kazimier Smith\thanks{Both authors contributed equally to this research.} \\
    Massachusetts Institute of Technology \\
    Cambridge, MA, USA \\
    \texttt{kazim119@mit.edu}
    \and
    Yucheng Lu\footnotemark[1]\\
    Harvard University \\
    Cambridge, MA, USA \\
    \texttt{yuchenglu@hks.harvard.edu}
    \and
    Qiaochu Fan \\
    New York University \\
    New York, NY, USA \\
    \texttt{qf2080@nyu.edu}
}

\date{}

\maketitle

\begin{abstract}
Public funding plays a central role in driving scientific discovery. To better understand the link between research inputs and outputs, we introduce FIND (Funding–Impact NSF Database), an open-access dataset that systematically links NSF grant proposals to their downstream research outputs, including publication metadata and abstracts. The primary contribution of this project is the creation of a large-scale, structured dataset that enables transparency, impact evaluation, and metascience research on the returns to public funding. To illustrate the potential of FIND, we present two proof-of-concept NLP applications. First, we analyze whether the language of grant proposals can predict the subsequent citation impact of funded research. Second, we leverage large language models to extract scientific claims from both proposals and resulting publications, allowing us to measure the extent to which funded projects deliver on their stated goals. Together, these applications highlight the utility of FIND for advancing metascience, informing funding policy, and enabling novel AI-driven analyses of the scientific process.
\end{abstract}

\noindent\textbf{Keywords:} NSF awards, research outputs, metascience, grant-publication linkage, natural language processing

\section{Introduction}
Public funding is a cornerstone of the scientific enterprise, shaping the trajectory of research across disciplines and driving advances with broad social and economic impact. Understanding the relationship between funded proposals and the research outputs they generate is important for scholars of science, policy makers, and funding agencies alike. To support such inquiry, we introduce the Funding–Impact NSF Database (FIND), a large-scale open resource that systematically links NSF grant proposals to their downstream publications.\footnote{Data available at \url{https://huggingface.co/datasets/kazim119/nsf_find}.} Unlike existing NSF award records, which often lack clean or consistent mappings to research outputs, FIND provides both structured metadata and textual content from proposals and publications, enabling researchers to examine funding outcomes with new precision. Beyond its value as a standalone dataset, FIND also opens the door to a variety of applications in metascience, economics, and natural language processing.

\section{Relevance}
NSF award metadata and their associated scientific publications offer useful data on the scientific process.
By linking award proposals to publications and citations, we construct a new resource that enables the study of science and science funding.
Our predictive models, including a hurdle approach with long-context language models, provide methodological advances for analyzing long-form text in a scientific context.

Our dataset also facilitates connections between potentially disparate scientific disciplines. Publications matched to NSF grants may not be in the same field, and our dataset makes clear the connection between the grant's field and the publication's field. Our scoring methodology could be used to show how publications in unexpected areas can achieve the original goals of a grant. We also hope our dataset can be used to improve the matching between publications and NSF grants awarded in the 20th century, for which the current data are poor. Such a matching would allow a detailed analysis of the evolution of American science over the past 75 years.

We begin with a combination of structured and unstructured data: the publicly available NSF grant database has clearly defined fields, but some of these fields contain mostly free-form text. Grant abstracts do not systematically list what the researchers intend to do with their funding. Publication abstracts do not include a point-by-point list of their findings. We use large language models to produce structured data from these unstructured fields by extracting specific research goals and results.

\section{Preliminaries and Background}
The National Science Foundation is a United States federal agency that funds scientific research in disciplines other than medicine \citep{nsf_about}. The National Institutes of Health support medical research \citep{nih_report_faqs}. The NIH has a robust dataset matching its funding to research outcomes via its RePORTER database, which directly links medical research grants to their subsequent publications and patents. The NSF lacks such a systematic matching. Although grants include a ``project outcomes report'' field, information is often missing or incomplete, and the data quality worsens for less recent grants. Moreover, the project outcomes report often does not include an explicit link (via DOI or otherwise) to publications resulting from the grant. Our dataset fills this gap.

The National Science Foundation funds research through \textit{grants} \citep{nsf_process}. Researchers seeking funding write a proposal in which they describe the research project, its goals, its methods, its potential outcomes, and the quantity and type of resources it requires. Proposals also describe how they will use the funds. Reviewers, often in groups and with expertise in the particular area of the proposal, examine proposals and choose which to accept \citep{hettich2006mining}. Researchers submit their proposals to one of eight \textit{directorates}, such as the Biological Sciences, which are broad research categories. Researchers may also list a more specific subfield for their proposal. Researchers can also submit proposals to renew funding for a specific project (if they need more time or more resources to complete it).

The NSF has evolved in structure since its inception in the 1950s. Here we describe some important changes relevant for our dataset (but certainly not all changes to the NSF in its history). First, Congress voted to enact the America COMPETES Act in 2007 and re-authorized it in 2010 \citep{competes2007}. This act substantially increased funding allocated to the National Science Foundation.

Second, in 2013, the Office of Science and Technology Policy required researchers funded by the NSF to upload data on their grants and resulting publications to a new database called the Public Access Repository. PAR aims to increase transparency of science funding in the wake of debates over what types of research should be funded, especially in the social sciences \citep{nsf2023glaciers}. Since PAR is relatively new, it is quite incomplete. We analyze it as part of our dataset construction and show that it matches only a small number of grants to their resulting publications.

\section{Related Work}
Perhaps the closest work to ours is \citet{rao2025nsfscifyminingnsfawards}. They collect publicly available data on NSF grants as we do and use large language models to extract ``scientific claims'' and ``investigation proposals'' from materials science grants. They intend their dataset as training data to fine-tune a smaller language model. We adapt their claim and proposal extraction for our purposes and apply it to all NSF grants after 2000. The extraction process is a vital part of our calculation of scientific success scores, but it could be costly. \citet{rao2025nsfscifyminingnsfawards} show, valuably, that smaller models can succeed at proposal extraction. This is an important takeaway for future research that might try to calculate scientific success scores on a larger dataset. Moreover, we link proposals from NSF grants to scientific findings in publications resulting from the grants. The linkage is one of our key contributions that allows us to evaluate the scientific success of grants. While the NSF award dataset\footnote{\url{https://www.nsf.gov/awardsearch/download.jsp}} does contain the field ``por'' (Project Outcomes Report for the General Public), for most of the entries this field is missing. Even in cases where it is populated, it rarely provides a clean linkage to research output (i.e. to publication DOI).

Another closely related work, \citet{jones2025leveragingglobalresearchinfrastructure}, compares the two main sources linking NSF awards to publications, Crossref \citep{crossref} and PAR \citep{nsfpar} and shows that Crossref is substantially more complete than PAR. We build on their work by (1) combining Crossref and PAR, (2) applying the linkage to all NSF awards from 2000 onwards, and (3) making our dataset publicly available. They approach the data from a metascience perspective, while we leverage the textual data to extract additional information.

\citet{jones2025leveragingglobalresearchinfrastructure} note that ``...the percentage of journal articles in Crossref with funder metadata ... averages less than 25\%.'' There are many articles that are not properly linked to their funding sources because they were issued prior to the NSF public access policy, because papers were published after the grant expired, or simply because principal investigators forgot to register their work. One avenue for future work is to use machine learning techniques similar to \citet{brown2023softsearchdatasetsstudyidentification} to create an extended linkage dataset.

More work exists on National Institutes of Health grants and related outcomes due to better data availability. While the NIH and the NSF fund different types of research, the ideas and conclusions in research on NIH funding are complementary to our work on the NSF. \citet{azoulay2019public} construct a novel dataset linking NIH grants to the biomedical patents they generate. They use the data to show that NIH funding boosts patent production; this conclusion demonstrates the value of linking funding sources to their outcomes. Other work has examined racial disparities in NIH funding \cite{ginther2018publications}. Further studies have looked at the effect of funding on postdoc career outcomes \cite{jacob2011postdoc}, scientific productivity \cite{jacob2011research}, and research novelty \cite{packalen2020nih}. Notably several of these studies use \textit{rejected} NIH grants as a baseline to estimate the impact of an accepted grant. Data on rejected NSF grants does not appear to be publicly available.

\citet{li2019nih} undertake a task very similar in spirit to our second NLP application. They examine keyword lists in NIH grants and resulting publications and use the lists to compute a measure of alignment between the grants and publications. We expand on this work in two ways: one, we improve the alignment metric by using a large language model's excellent ability to process unstructured text. Two, we cover many more disciplines than the NIH grant-publication matching which is largely specialized to biomedical research. There are other sources of federal research funds, like the Department of Energy, and future work could attempt to match grants from these other agencies to resulting publications and patents. A full dataset linking all federally funded research to its outcomes would be extremely valuable.

Finally, firms like DataSeer (https://dataseer.ai) also contribute to standardizing and quantifying open science. Their tools can measure, for example, the degree to which publications in a given field generate open source code. They note that research funders receive ``a better return on investment'' when resulting publications generate open access data and code. Critically, DataSeer does not link funding sources to publications, so they cannot directly measure the amount of open access data and code resulting from a particular NSF grant. Our dataset fills the gap and we hope it can be used to study the impact of federal funding on open access science.

\begin{table*}[htbp]
\centering
\caption{Comparison of Related Work}
\label{tab:literature}
\begin{tabular}{lccc}
\hline
\textbf{Study} & \textbf{Dataset} & \textbf{Method} & \textbf{Limitation} \\
\hline
Rao et al. (2025) & NSF Grants & LLM & No link to outcomes \\
Jones and Haberman (2025) & NSF Grants & Database & No text analysis \\
\textbf{Our Work} & \textbf{NSF Grants} & \textbf{Database + LLM} & \textbf{Focus on post-2000 NSF grants} \\
\hline
\end{tabular}
\end{table*}

\section{Methods}

\subsection{Initial data collection}
We download our raw data from the NSF Award Search database. The NSF provides a web interface where users can search for awards based on award number, keywords, authors, etc. They also provide JSON files containing data on all NSF awards in each year from 1960-2025. The database includes historical records before 1960, but we focus on the data from 1960 onwards to facilitate linking with scientific papers. Matching older NSF grants to their scientific output is a valuable avenue for future research on the history of science in the United States.

\begin{figure}[htbp]
    \centering
    \includegraphics[width=0.9\linewidth]{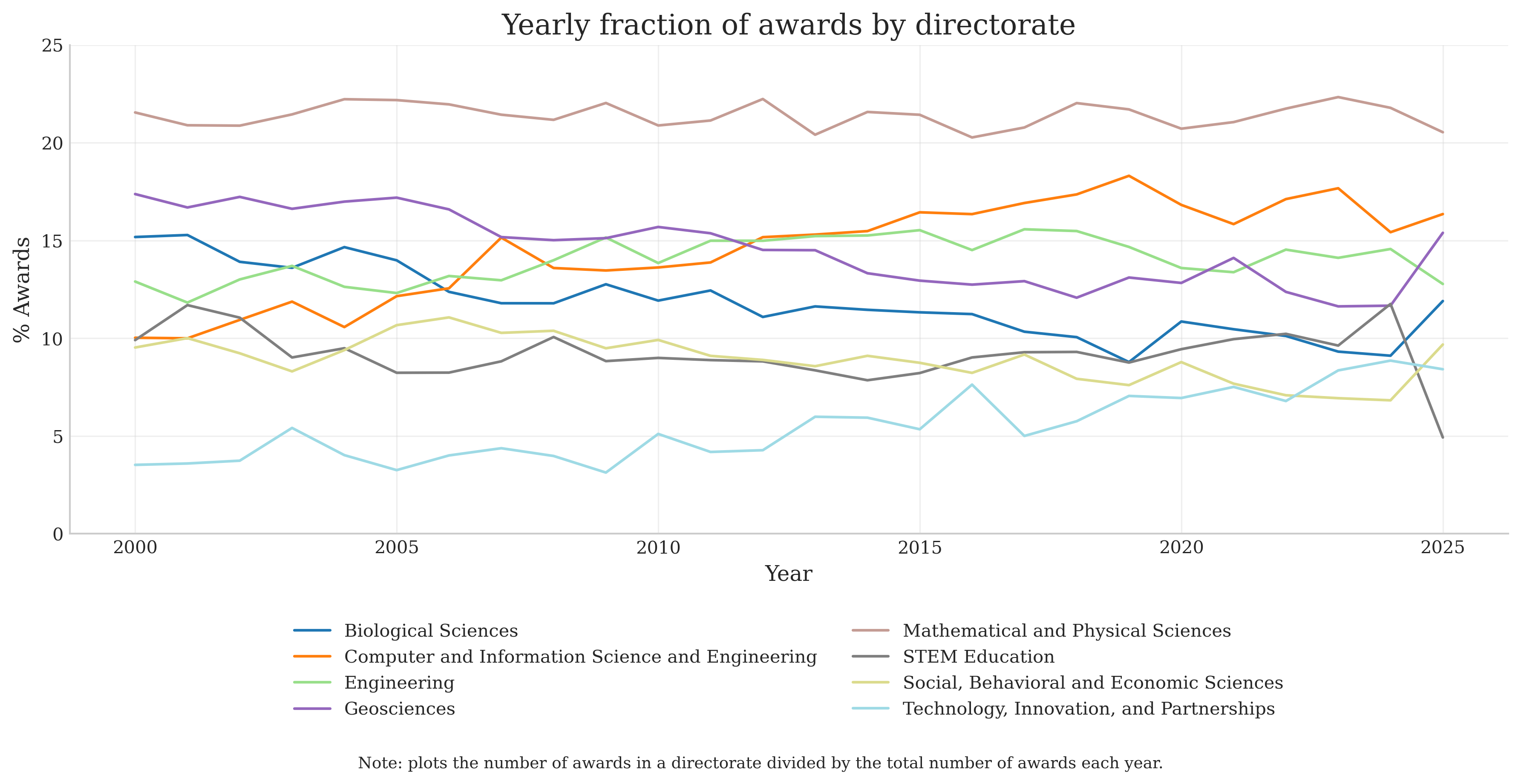}
    \caption{Grant allocation by directorate}
    \label{fig:frac_awards_directorates_time}
\end{figure}

\begin{figure}[htbp]
    \centering
    \includegraphics[width=0.9\linewidth]{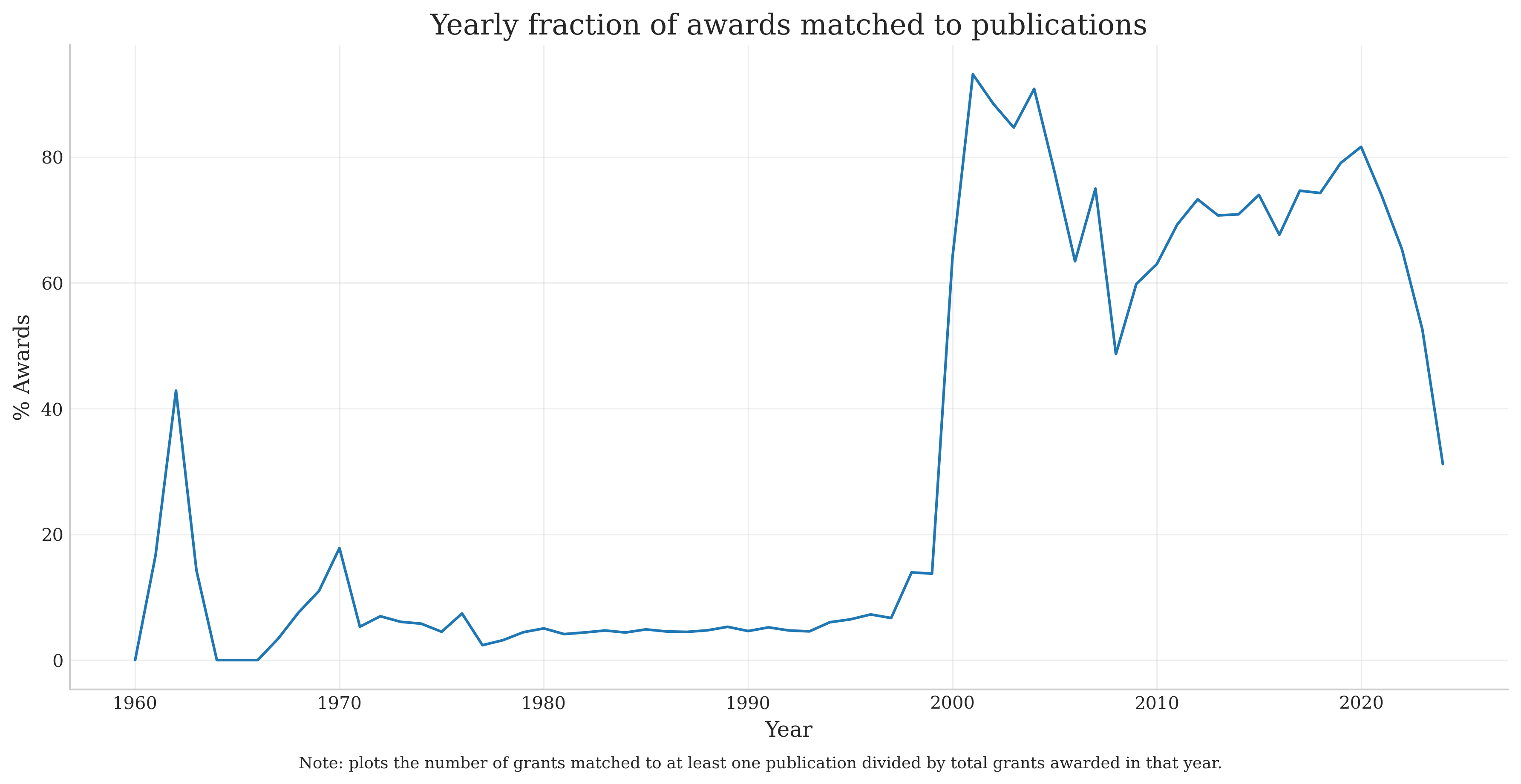}
    \caption{Grant-publication match rate}
    \label{fig:awd_time}
\end{figure}

We process the raw JSON files to obtain a dataset of 525,927 unique award IDs. The NSF assigns each award to a broad category called a ``directorate''. A small number of awards are assigned to administrative directorates such as the Office of the Director. We largely ignore these awards in our analysis since they are likely part of the NSF administration rather than true scientific production. After dropping these awards, Figure \ref{fig:frac_awards_directorates_time} plots the fraction of award IDs in each directorate over time for the universe of awards from 2000-2025. We focus on this period because all directorates exist after 2000 and because the match between NSF grants and publications is far better after 2000 (Figure \ref{fig:awd_time}). While the fraction of grants in each directorate remains relatively steady over time, the share of Technology, Innovation, and Partnerships grants more than doubles from about 3\% of all grants in 2000 to more than 7\% in 2024.

\begin{figure}[htbp]
    \centering
    \includegraphics[width=0.9\linewidth]{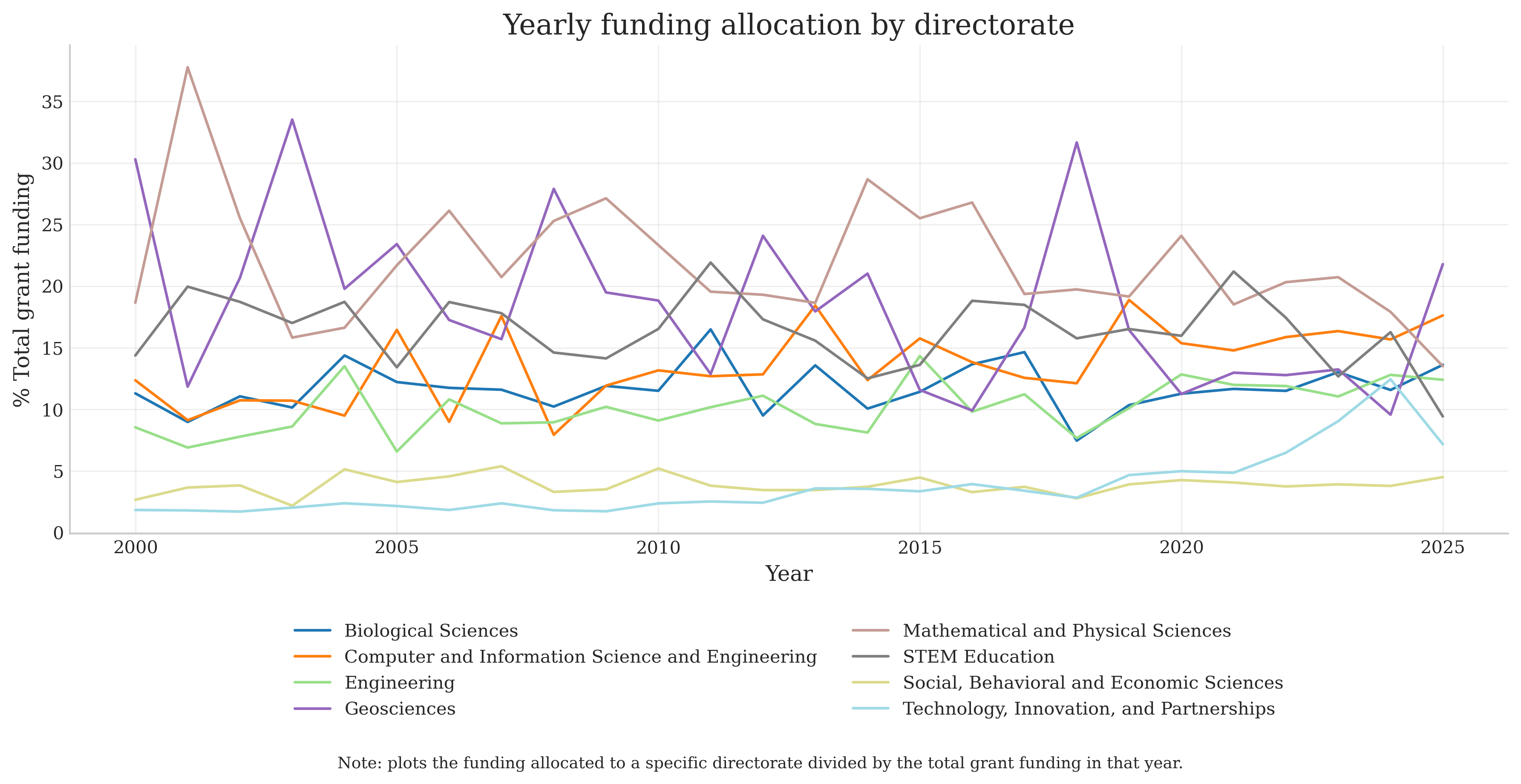}
    \caption{Fraction of grant funding in each directorate over time}
    \label{fig:frac_funding_directorates_time}
\end{figure}

Note that Figure \ref{fig:frac_awards_directorates_time} ignores the total monetary amount of the award. Figure \ref{fig:frac_funding_directorates_time} shows the fraction of total grant funding allocated to each directorate in each year. The social sciences receive substantially less funding. Research costs in the social sciences are typically lower, so more work is needed to determine whether the social sciences are systematically underfunded.

The JSON files from the NSF award database contain 800 unique metadata fields for each award, although many of these fields are empty for most awards. For example, ``pgm\_ele\_0\_pgm\_ele\_name'' contains more granular information on the scientific field to which the grant contributes. The ``por'' field contains the award's Project Outcomes Report, which is a required \citep{competes2007} report documenting the scientific contributions that resulted from the NSF award.

\subsection{Matching awards to scientific publications}
The next step in the dataset construction is finding the scientific publications associated with each NSF grant. We primarily use Crossref \citep{crossref} for this task. Crossref is a nonprofit organization that maintains a database of metadata about academic publications. Publishers enter information about their articles in the database. For this step, the most important piece of metadata in Crossref is funding sources for the publication. When an NSF grant supported a publication, the publisher can enter the NSF award ID into Crossref's database. This allows us to link NSF grants to all the publications they supported. While participation in Crossref is voluntary, we find that 35\% of NSF grants in our dataset (1960-2025) have at least one publication associated with them in Crossref. Although this is far from full coverage, it still provides a large dataset for downstream analysis and model finetuning.

We also use the NSF's official Public Access Repository \citep{nsfpar} to attempt to find publications that might not appear in Crossref. PAR began after a 2013 law required the NSF to disclose more information to the public about its grants and their outcomes. PAR is ``the designated repository where NSF-funded investigators deposit peer-reviewed, published journal articles'' \citep{nsfpar}. Although submitting articles to PAR became mandatory in 2016, adherence appears somewhat poor: we find that only 43.6\% of NSF award IDs from 2017 onwards have a publication listed in PAR. From 2000-2016, this drops to 0.05\%. Finally, 25\% of publications found in PAR are also found in Crossref. We hope our publicly available dataset will help further the NSF's goal of increasing public awareness of science funding.

Practically, we create a list of publications associated with an NSF award ID by taking the union of publications listed in Crossref and publications listed in PAR. Of our 525,927 total award IDs, 224,524 of them are matched to at least one publication (42.7\%). Figure \ref{fig:awd_time} shows this fraction over time. Coverage improves significantly after 2000 likely because the Digital Object Identifier system was introduced in that year \cite{what_is_doi}. Improving the linkage between pre-2000 NSF grants and their publication outcomes is another important task for the study of the history of American science, and we hope our dataset can contribute to that effort.

\begin{figure}[htbp]
    \centering
    \includegraphics[width=0.9\linewidth]{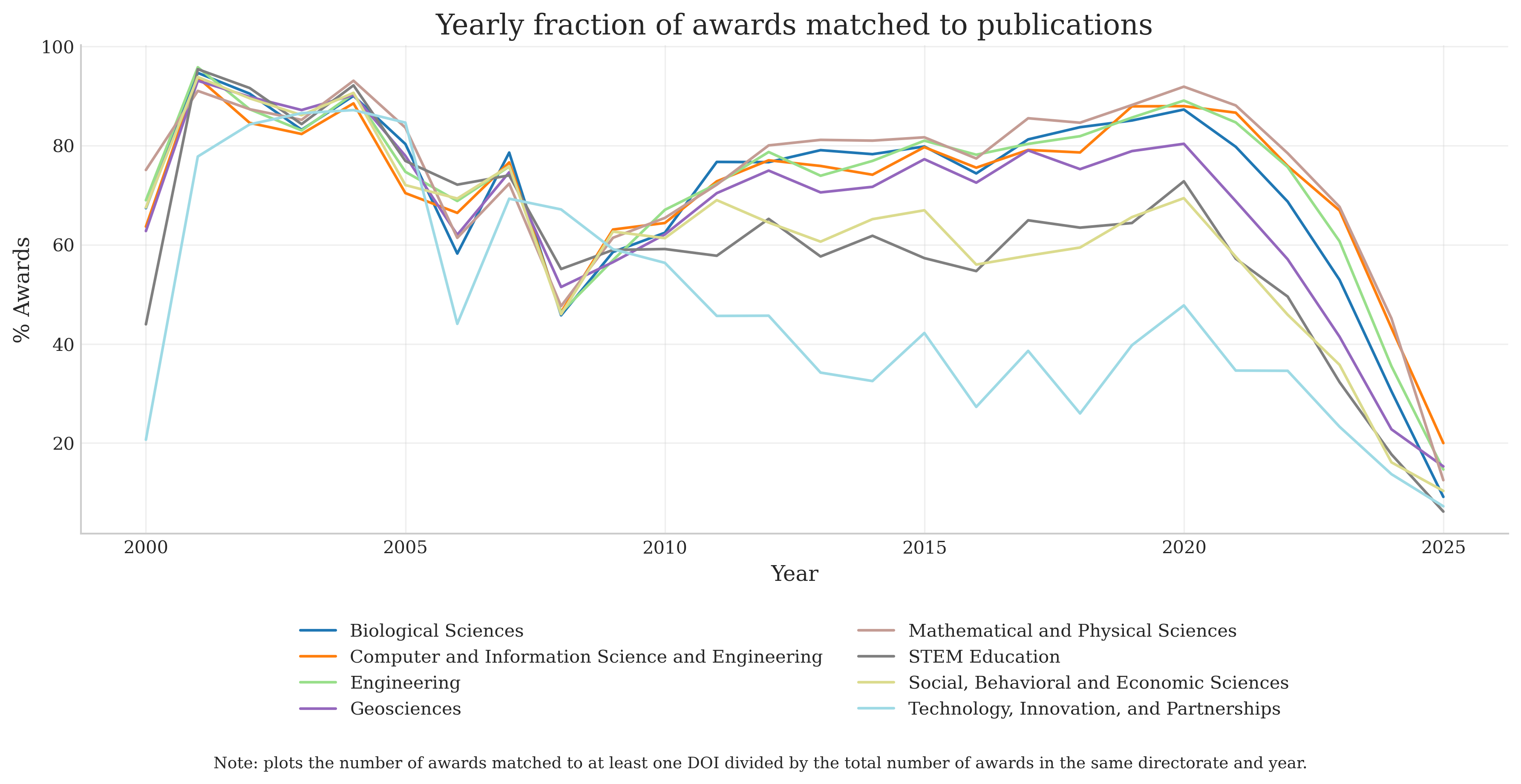}
    \caption{Grant-publication match rate by directorate}
    \label{fig:frac_awards_within_directorates}
\end{figure}

Figure \ref{fig:frac_awards_within_directorates}  shows the data coverage from 2000 onwards broken down by directorate. The Social Sciences and STEM education have the lowest match rates, and it is worth noting that their match rates decrease after about 2008 while the Mathematical Sciences, Computer Sciences, and Engineering increase. We hope our dataset can help researchers explore the economic and political forces behind this divergence. Practically, using our dataset to finetune language models might lead to worse performance on social science related grants because we have fewer data points in that category.

\subsection{Matching scientific publications to further metadata}
The final step is collecting additional metadata about each publication that results from one of the NSF grants in our data. We primarily use Crossref \citep{crossref} and OpenAlex \citep{priem2022openalex} for this. Crossref only uses publications registered with its database to calculate citation counts, so it likely provides an underestimate of the true citation count. To remedy this, we merge in citation information from OpenAlex. OpenAlex is the successor to Microsoft Academic Graph and aims to provide a completely open database of scholarly work. It includes data from Crossref and other sources (e.g. PubMed). OpenAlex contains an abstract for 512,644 publications in our dataset and a citation count for 936,703 publications. We restrict our first natural language processing task, predicting citation counts, to the subset of publications with a citation count available in OpenAlex. Our second task relies on publication abstracts, so we restrict it to that subsample of the OpenAlex data.

\begin{figure}[htbp]
    \centering
    \includegraphics[width=0.9\linewidth]{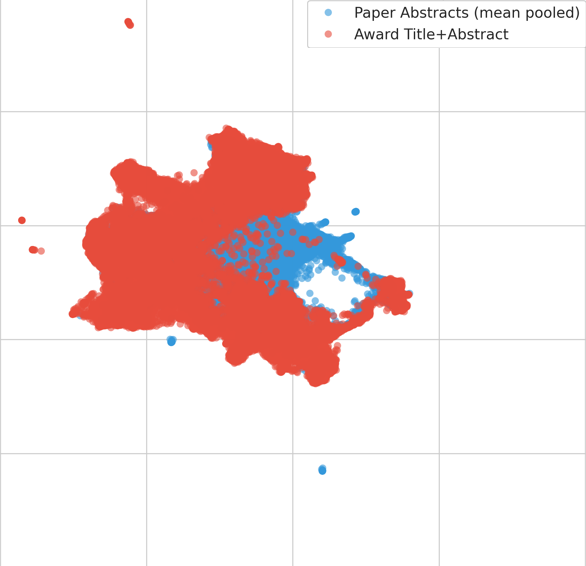}
    \caption{Award and Publication Embedding Comparison via UMAP dimension reduction}
    \label{fig:umap_comparison}
\end{figure}

Figure \ref{fig:umap_comparison} compares the semantic embedding spaces of award abstracts and paper abstracts using the SPECTER2 \citep{Singh2022SciRepEvalAM} model. Each point represents a document embedding projected into two dimensions via UMAP \citep{mcinnes2020umapuniformmanifoldapproximation} for visualization. Award-level embeddings (red) were computed from the concatenation of award titles and abstracts, while paper-level embeddings (blue) were obtained by mean-pooling the SPECTER2 embeddings of individual paper abstracts linked to the corresponding awards. The visualization reveals substantial overlap between the two distributions, indicating that award descriptions capture much of the same topical structure as the resulting papers, despite being broader and more heterogeneous in scope. A small region to the top right shows a clean separation of paper abstracts from award abstracts, suggesting topical or linguistic divergence for a subset of research outputs.

\section{Experiments}

\subsection{Predicting citations}
\paragraph{Hurdle Model} To account for the highly skewed and zero-inflated distribution of citation counts across publications linked to NSF awards, we implement a \textbf{two-stage hurdle modeling approach}. This framework separately models (1) the likelihood that an award leads to any cited research, and (2) the expected citation impact conditional on at least one cited publication. The hurdle model framework allows us to disentangle the drivers of ``having any impact'' from those that predict ``how much impact'' a grant might have. For example, broader or applied proposals may be more likely to produce at least one publication, while highly technical or ambitious proposals may produce fewer but highly cited outputs. This decomposition provides a more interpretable and flexible framework for assessing the predictive value of grant proposal content and metadata.

\paragraph{Data and Features} We analyze NSF awards from 2000-2020, aggregating citation counts to the award level. Our approach enforces strict ex-ante prediction using only metadata available at funding time: award amount, temporal information (effective/expiration dates, duration), organizational directorate, and textual content from titles and abstracts. We generate dense embeddings from award text using SPECTER2 \citep{Singh2022SciRepEvalAM}, a powerful science-domain adapted embedding model. Structured features undergo robust scaling and target encoding, with optional polynomial interaction terms to capture non-linearities.
We first aggregate our dataset to the \emph{award level}, computing the \emph{average citation count} across all publications associated with each NSF grant. This transformation ensures each data point represents a distinct research award, capturing its total downstream citation impact rather than individual paper-level variation. We partition data into training (60\%), validation (20\%), and test (20\%) sets with stratified sampling.

\paragraph{Model Architecture} In Stage 1, we train binary classifiers (LightGBM \citep{10.5555/3294996.3295074}, MLPs, elastic-net logistic regression) to predict whether an award receives any citations, optimizing hyperparameters via Optuna with 5-fold cross-validation and ROC-AUC as the objective. In Stage 2, the target becomes the log-transformed citation counts for awards with citations. Beyond individual models, we construct voting, stacking, and calibrated weighted (only for classification) ensembles to leverage complementary predictive strengths.

\begin{figure}[htbp]
    \centering
    \begin{subfigure}{0.48\textwidth}
        \centering
        \includegraphics[width=\textwidth]{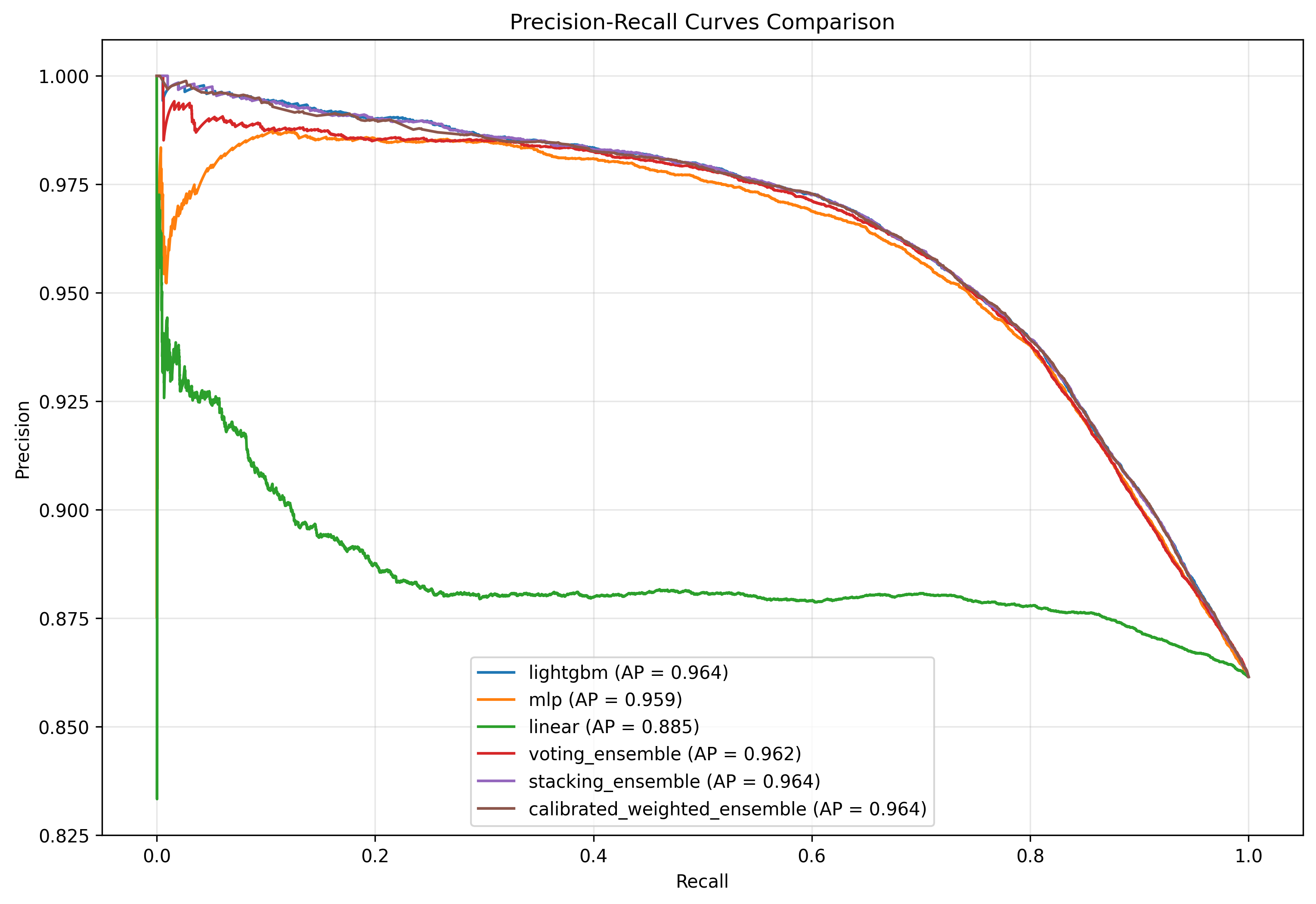}
        \caption{Precision Recall Curve}
        \label{fig:plot1}
    \end{subfigure}
    \hfill
    \begin{subfigure}{0.48\textwidth}
        \centering
        \includegraphics[width=\textwidth]{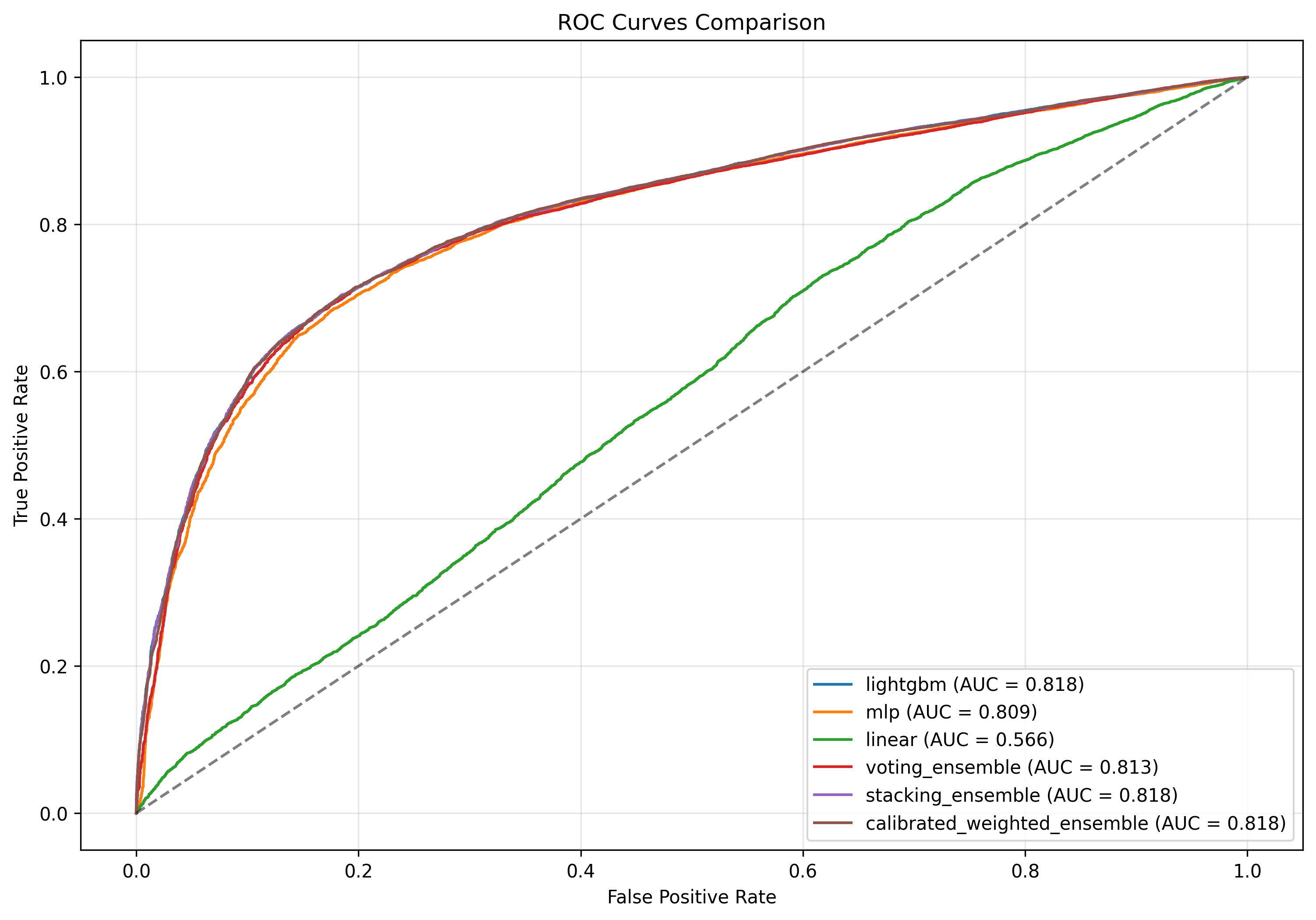}
        \caption{ROC Curve}
        \label{fig:plot2}
    \end{subfigure}
    \caption{First Stage Performance}
    \label{fig:both_plots}
\end{figure}

\paragraph{Results} Figure \ref{fig:both_plots} presents the ROC and precision-recall curves for binary classification of citation occurrence. In the first stage classification, the models achieve strong discriminative performance, with AUC scores ranging from 0.809 to 0.818 for gradient boosting and neural network approaches, and average precision scores exceeding 0.96. The precision-recall curves demonstrate that models maintain precision above 0.95 across a wide range of recall values, indicating robust identification of awards likely to generate cited publications. This high level of performance suggests that award metadata and textual content at the time of funding contain substantial signal about whether a project will produce cited research outputs. Incidentally, LightGBM (AUC = 0.818, AP = 0.964) alone seems to deliver most of the performance. 

\begin{figure}[htbp]
    \centering
    \includegraphics[width=0.9\linewidth]{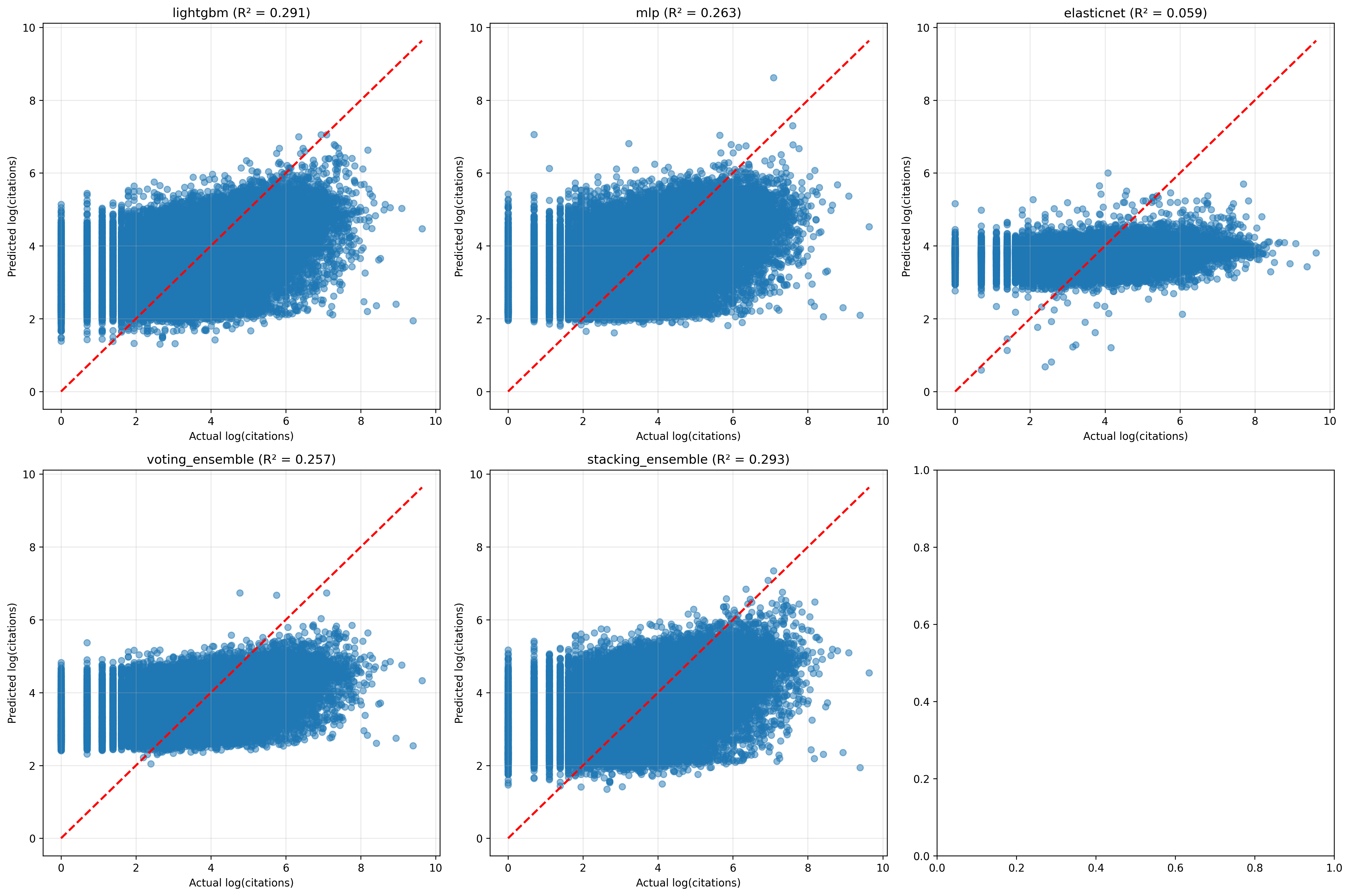}
    \caption{Second Stage Performance}
    \label{fig:secondstage}
\end{figure}

Figure \ref{fig:secondstage} shows the performance of the second stage regression. While the predictive performance is not particularly strong, it is nonetheless nontrivial, especially considering that our model uses award-level information rather than features derived from the actual papers themselves. Predicting citation counts is inherently difficult, as citations are influenced by numerous unobserved social, disciplinary, and temporal factors. Interestingly, even studies that leverage features from research papers themselves typically report values in the range of 0.4-0.5 (e.g., \citep{petersen2011statistical,10.1007/978-3-319-07467-2_12}), suggesting that our approach captures a meaningful portion of the variation despite relying on higher-level award metadata.

\subsection{Grant success scores}
NSF officials and scientists are likely interested in how ``successful'' grants are. Success has several possible definitions in this context; we focus on publication outcomes given the linking we create. NSF grants aim to advance scientific discovery, and their results are disseminated in peer-reviewed publications. Although many of the grants in our dataset result in at least one publication, it is not obvious that these publications achieve the grant's original goals. To measure this, we calculate a ``scientific success score'' for each grant (we sometimes refer to it as the ``alignment score''). This score measures how well each publication achieves the research proposals in the grant. A grant might produce many publications, but if few of them align with the grant's original aims, the grant has not been entirely successful. On the other hand, a grant that produces a small number of publications which directly achieve its goals could be considered more successful.

To construct this score, we take advantage of large language models' ability to quickly extract information from unstructured text. First, we use gpt-4.1-nano to process NSF grant abstracts. We ask the model to extract ``verifiable claims'' and ``investigation proposals'' (see Appendix A for the full prompt) from each grant abstract. A verifiable claim is an existing fact which the grant states. A grant researching plant growth might note that ``plants require sunlight to produce energy'' before describing its research goals. An investigation proposal is one such goal: some scientific project the grant applicant will carry out using the funding. The grant on plant research might state ``we will measure the impact of partial shade on photosynthesis.'' Grants can have multiple proposals. From 525,927 grants we extract 2,032,600 unique investigation proposals; the median grant has five proposals. We fail to extract proposals from 40,877 grants. The pragmatic reason to extract verifiable claims is to reduce the number of instances in which the model interprets settled science as proposed research \cite{rao2025nsfscifyminingnsfawards}. While we do not use verifiable claims in our NLP applications here, we include them in the dataset for future research. For example, the degree to which a grant relies on existing facts might proxy for its novelty; this is a useful measure for studying changing appetites for risk at the NSF and among scientists.

Next, we process the abstracts of the 512,644 publications matched to our grants data again using gpt-4.1-nano. In particular we extract ``reported methods'' and ``scientific findings'' (see Appendix A for the full prompt). Reported methods are scientific techniques, for example, electrochemical impedance spectroscopy, that the authors used to conduct their research. Scientific findings describe the results of the publication. For example, the publication using electrochemical impedance spectroscopy might find that ``A new lithium-sulfur cathode design improves cycling stability by 40\% over conventional designs.'' We focus on scientific findings here but include reported methods in the dataset for future work. Again, asking the model to extract reported methods helps delineate true research results from techniques and methodologies. We extract at least one scientific finding from 472,861 of 512,643 publications; among these, the median publication has three scientific findings.

Each grant has one or more investigation proposals, and each publication has one or more scientific findings. Each grant is matched to one or more publications. Within each grant, we generate all pairwise combinations of a proposal and a finding. This generates 12,424,084 (proposal, finding) pairs; each of these is associated with a unique NSF grant. Our goal is to assign a score to each of these pairs which captures how well the scientific finding achieves the goals set out in the grant proposal. Because proposals and findings are free-form, it is not obvious how to assign a score with a traditional NLP algorithm. Instead, we use gpt-4.1-nano to generate the scores: we pass each pair to the model and ask it to assign a score from zero to 100 capturing how well the finding achieves the goals in the proposal (see Appendix A for the full prompt). We are able to assign a score to 12,424,016 of the pairs, and the median pair has a score of 20. Since we generate all pairwise combinations of proposals and findings, it is unsurprising that the median score is low. Some pairs will be mismatched since a given publication will not necessarily attempt to answer all the questions in the original grant (indeed, since the median grant generates two publications, it seems plausible that grant awardees publish separate papers for distinct proposals in the grant). The mean pair has a score of 30.81 and the standard deviation of scores is 25.44. Aggregating to the grant level, the median grant has five (proposal, finding) pairs with a score above 50.

\begin{figure}[htbp]
    \centering
    \includegraphics[width=0.9\linewidth]{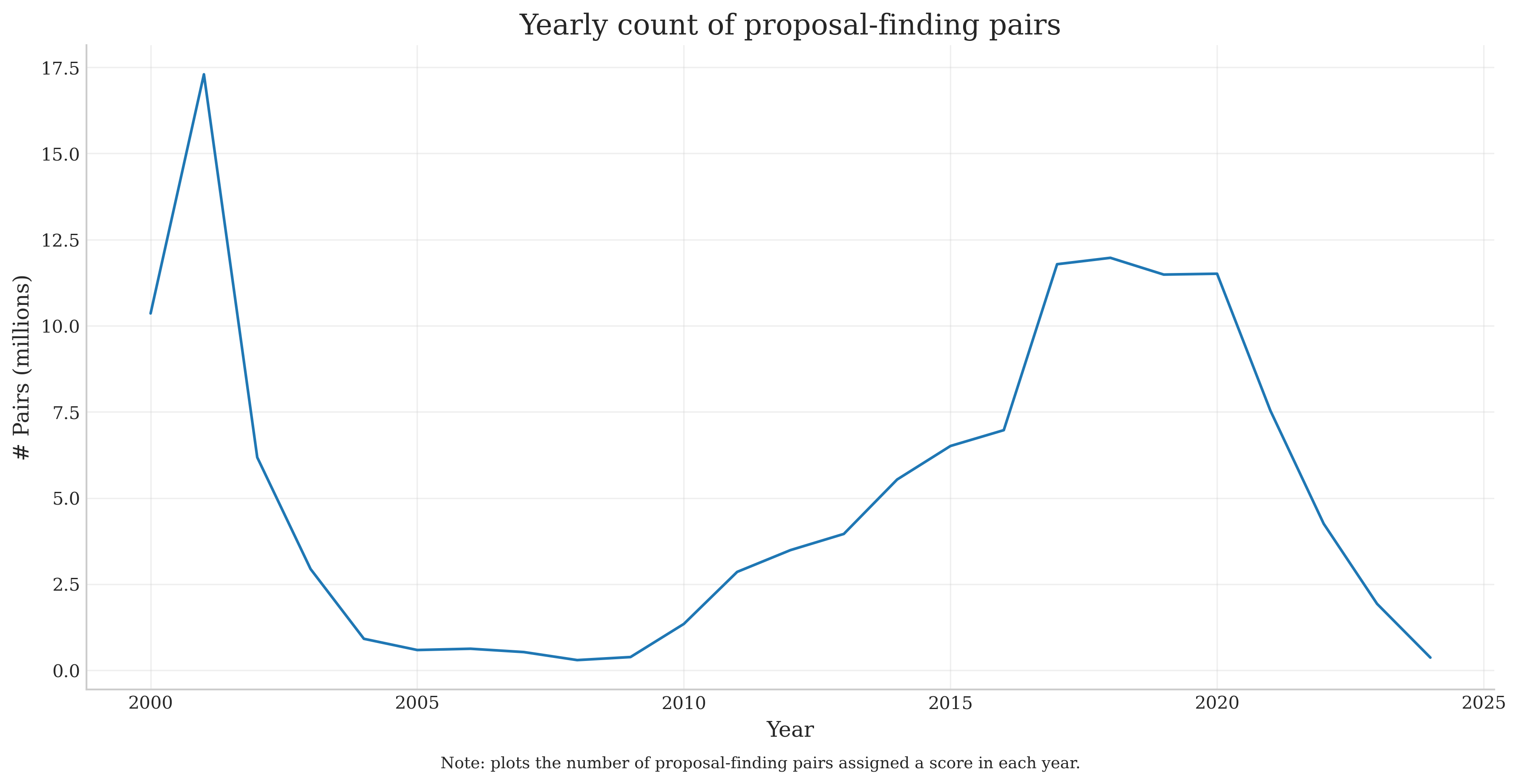}
    \caption{Number of proposal-finding pairs with a score over time}
    \label{fig:score_count_year}
\end{figure}

\begin{figure}[htbp]
    \centering
    \includegraphics[width=0.9\linewidth]{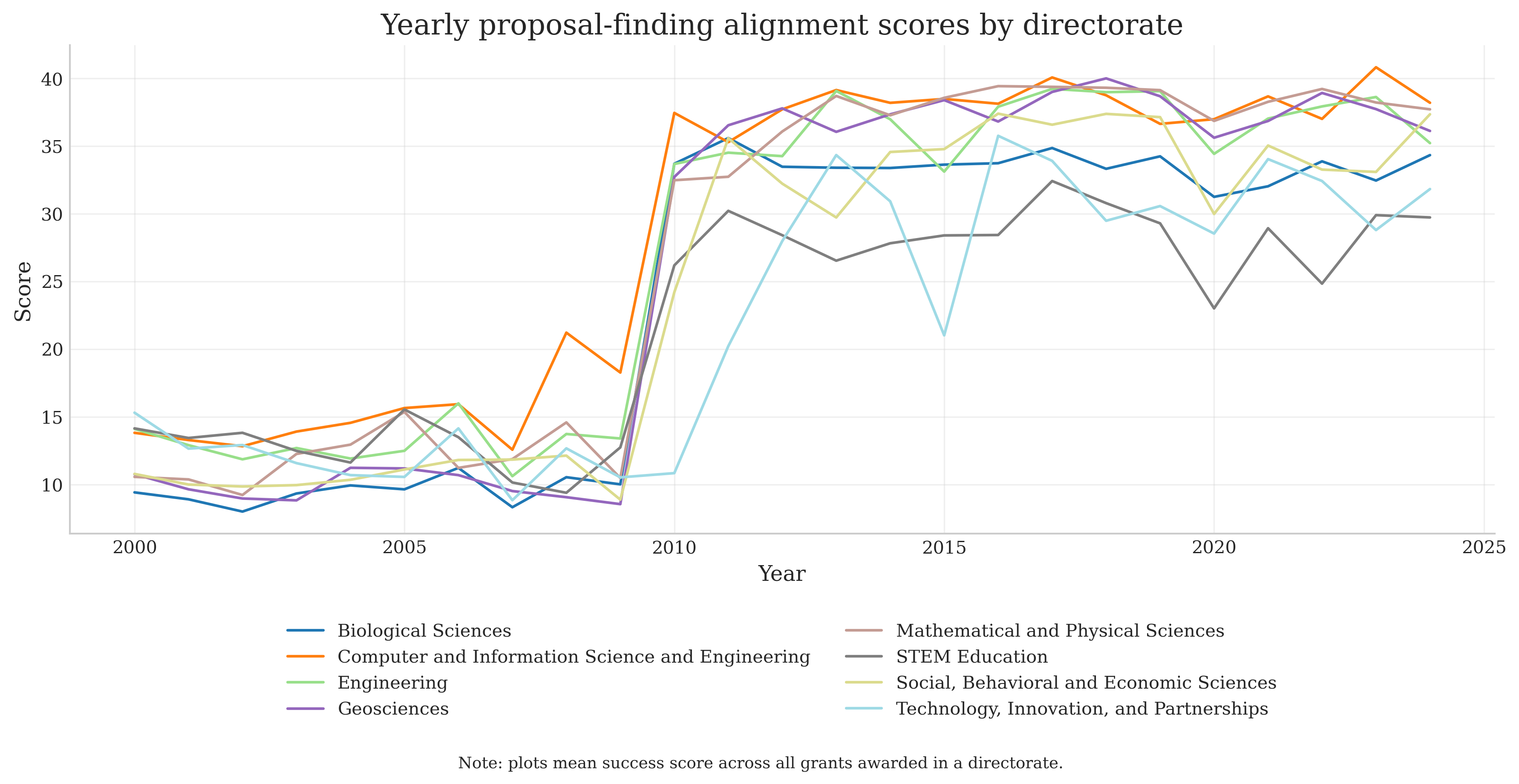}
    \caption{Scores over time by directorate}
    \label{fig:score_directorate_time}
\end{figure}

Finally, we investigate the pattern of these scientific success scores over time. Within each grant award year (the year in which the grant began, not the expiration year), we calculate the mean success score across all (proposal, finding) pairs.
\begin{figure}[htbp]
    \centering
    \includegraphics[width=0.9\linewidth]{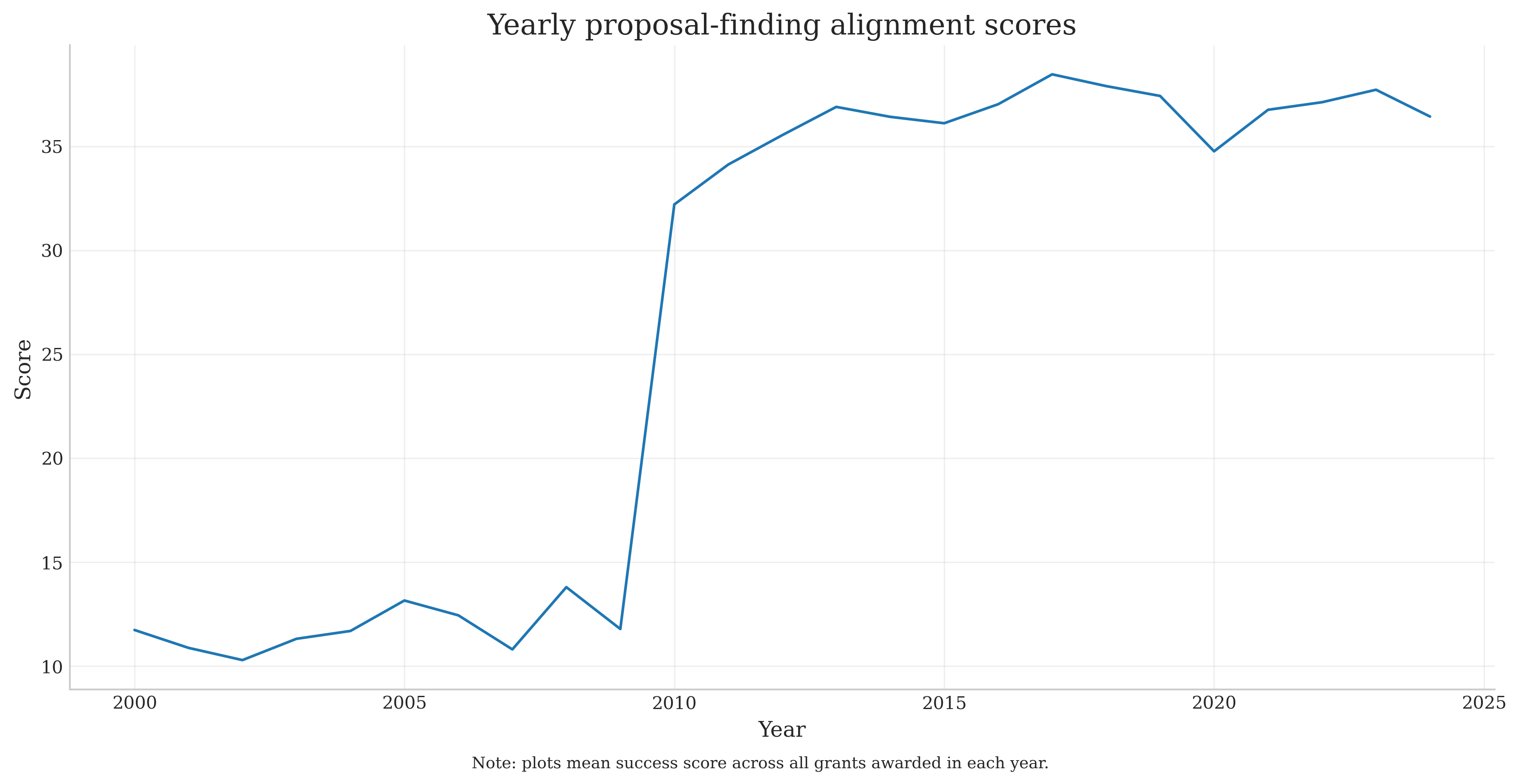}
    \caption{Scores over time}
    \label{fig:score_mean_year}
\end{figure}
Figure \ref{fig:score_mean_year} plots these means from 2000 to 2024 and shows a sharp increase in the average score starting in 2010. Figure \ref{fig:score_count_year} shows that this did not result purely from a decrease in the number of scored pairs, and Figure \ref{fig:score_directorate_time} shows that the pattern persists across NSF directorates. Our primary suggestion for future work is to investigate this drastic change. A change in the way the NSF awards or evaluates grants could generate the sudden jump; since there is no clear upward trend, a policy change is a plausible explanation. Two notable policy changes around 2010 were:
\begin{enumerate}
\item The federal government increased NSF grant funding as part of the America COMPETES Act \cite{competes2007} which was reauthorized in 2010. If this funding improved scientists' ability to carry out the research proposed in their grants, it might have led to more publications that directly achieve grant proposals.
\item The NSF began requiring quarterly reports on grant progress. This might have simply increased pressure on scientists to deliver on their proposed research, thereby increasing success scores. Alternatively, pressure to deliver positive reports might have led scientists to alter the language they use in grant proposals and in resulting publications. Since our methodology relies heavily on the actual language researchers use rather than the true underlying facts, a change in language could cause a sharp change in calculated success scores.
\end{enumerate}

\section{Conclusion}

We present the FIND dataset, an open-access collection of National Science Foundation grants across all disciplines linked to the scientific publications they ultimately generate. We demonstrate its utility through two NLP applications: predicting citation counts and calculating grant ``success scores''.

We suggest two directions for future work. First, filling in the matching between pre-2000 NSF grants and their resulting publications would create a valuable dataset for metascience research. This is a difficult task that would likely involve multiple methods and data sources. A first step might be to research the methods Crossref uses to link publications to grants and to try to extend those methods to unmatched publications. We hope our dataset can serve as a baseline to validate this matching. In particular, our (proposal, finding) pair alignment scores offer an estimate of how well publication and grant text should match. Significant divergence between pre-2000 pair scores and post-2000 pair scores could indicate a matching issue.

Second, the substantial increase in alignment scores after 2009 demands further inquiry. Both changes in NSF funding and changes in reporting requirements may have contributed, and future work could disentangle and quantify these two forces. \citet{packalen2020nih} also note that ``...NIH funding has become more conservative despite initiatives to increase funding for innovative projects.'' If innovative research is less predictable from the grants that fund it, then innovative publications will have lower alignment scores with their funding grants. An increase in risk-aversion at the NSF after 2009 could therefore explain the sudden increase in alignment scores.

\clearpage
\pagebreak
\bibliographystyle{plainnat}
\bibliography{ref} %

@misc{rao2025nsfscifyminingnsfawards,
      title={NSF-SciFy: Mining the NSF Awards Database for Scientific Claims}, 
      author={Delip Rao and Weiqiu You and Eric Wong and Chris Callison-Burch},
      year={2025},
      eprint={2503.08600},
      archivePrefix={arXiv},
      primaryClass={cs.CL},
      url={https://arxiv.org/abs/2503.08600}, 
}

@misc{brown2023softsearchdatasetsstudyidentification,
      title={Soft-Search: Two Datasets to Study the Identification and Production of Research Software}, 
      author={Eva Maxfield Brown and Lindsey Schwartz and Richard Lewei Huang and Nicholas Weber},
      year={2023},
      eprint={2302.14177},
      archivePrefix={arXiv},
      primaryClass={cs.DL},
      url={https://arxiv.org/abs/2302.14177}, 
}

@misc{jones2025leveragingglobalresearchinfrastructure,
      title={Leveraging the Global Research Infrastructure to Characterize the Impact of National Science Foundation Research}, 
      author={Jamaica Jones and Ted Habermann},
      year={2025},
      eprint={2501.06843},
      archivePrefix={arXiv},
      primaryClass={cs.DL},
      url={https://arxiv.org/abs/2501.06843}, 
}

@article{azoulay2019public,
  title={Public R\&D investments and private-sector patenting: evidence from NIH funding rules},
  author={Azoulay, Pierre and Graff Zivin, Joshua S and Li, Danielle and Sampat, Bhaven N},
  journal={The Review of economic studies},
  volume={86},
  number={1},
  pages={117--152},
  year={2019},
  publisher={Oxford University Press}
}

@inproceedings{hettich2006mining,
  title={Mining for proposal reviewers: lessons learned at the national science foundation},
  author={Hettich, Seth and Pazzani, Michael J},
  booktitle={Proceedings of the 12th ACM SIGKDD international conference on Knowledge discovery and data mining},
  pages={862--871},
  year={2006}
}

@article{li2019nih,
  title={Are NIH-funded publications fulfilling the proposed research? An examination of concept-matchedness between NIH research grants and their supported publications},
  author={Li, Kai and Yan, Erjia},
  journal={Journal of Informetrics},
  volume={13},
  number={1},
  pages={226--237},
  year={2019},
  publisher={Elsevier}
}

@article{ginther2018publications,
  title={Publications as predictors of racial and ethnic differences in NIH research awards},
  author={Ginther, Donna K and Basner, Jodi and Jensen, Unni and Schnell, Joshua and Kington, Raynard and Schaffer, Walter T},
  journal={PloS one},
  volume={13},
  number={11},
  pages={e0205929},
  year={2018},
  publisher={Public Library of Science San Francisco, CA USA}
}

@article{jacob2011postdoc,
  title={The impact of NIH postdoctoral training grants on scientific productivity},
  author={Jacob, Brian A and Lefgren, Lars},
  journal={Research policy},
  volume={40},
  number={6},
  pages={864--874},
  year={2011},
  publisher={Elsevier}
}

@article{jacob2011research,
  title={The impact of research grant funding on scientific productivity},
  author={Jacob, Brian A and Lefgren, Lars},
  journal={Journal of public economics},
  volume={95},
  number={9-10},
  pages={1168--1177},
  year={2011},
  publisher={Elsevier}
}

@article{packalen2020nih,
  title={NIH funding and the pursuit of edge science},
  author={Packalen, Mikko and Bhattacharya, Jay},
  journal={Proceedings of the National Academy of Sciences},
  volume={117},
  number={22},
  pages={12011--12016},
  year={2020},
  publisher={National Academy of Sciences}
}

@misc{nih_report_faqs,
  title = {Frequently Asked Questions (FAQs)},
  author = {{National Institutes of Health}},
  howpublished = {\url{https://report.nih.gov/faqs}},
  note = {Accessed: 2025-10-06}
}

@misc{nsf_about,
  title = {About {NSF}},
  author = {{National Science Foundation}},
  howpublished = {\url{https://www.nsf.gov/about}},
  note = {Accessed: 2025-10-06}
}

@misc{nsf_process,
  title = {Overview of the NSF Proposal and Award Process},
  author = {{National Science Foundation}},
  howpublished = {\url{https://www.nsf.gov/funding/overview}},
  note = {Accessed: 2025-10-06}
}

@misc{competes2007,
  title = {America {COMPETES} {Act}},
  author = {{110th Congress}},
  howpublished = {Public Law 110-69},
  year = {2007},
  month = aug,
  day = {9},
  note = {121 Stat. 572},
  url = {https://www.govinfo.gov/app/details/PLAW-110publ69}
}

@techreport{nsf2023glaciers,
  title = {Response to Media Reports on {NSF}-supported Glaciers and Glaciology Award},
  author = {{National Science Foundation}},
  institution = {National Science Foundation},
  year = {2023},
  month = aug,
  url = {https://nsf-gov-resources.nsf.gov/2023-08/NSF_Response_supported_Glaciers_and_Glaciology_Award.pdf}
}

@misc{crossref,
  title = {Crossref},
  author = {{Crossref}},
  howpublished = {\url{https://www.crossref.org/}},
  year = {2025}
}

@misc{nsfpar,
  title = {{NSF} Public Access Repository ({NSF-PAR})},
  author = {{National Science Foundation}},
  howpublished = {\url{https://par.nsf.gov/}},
  year = {2025}
}

@misc{what_is_doi,
  title = {What is a DOI?},
  author = {{doi Foundation}},
  howpublished = {\url{https://www.doi.org/the-identifier/what-is-a-doi/}},
  note = {Accessed: 2025-10-06}
}

@article{priem2022openalex,
  title={OpenAlex: A fully-open index of scholarly works, authors, venues, institutions, and concepts},
  author={Priem, Jason and Piwowar, Heather and Orr, Richard},
  journal={arXiv preprint arXiv:2205.01833},
  year={2022}
}

@inproceedings{Singh2022SciRepEvalAM,
  title={SciRepEval: A Multi-Format Benchmark for Scientific Document Representations},
  author={Amanpreet Singh and Mike D'Arcy and Arman Cohan and Doug Downey and Sergey Feldman},
  booktitle={Conference on Empirical Methods in Natural Language Processing},
  year={2022},
  url={https://api.semanticscholar.org/CorpusID:254018137}
}

@inproceedings{10.5555/3294996.3295074,
author = {Ke, Guolin and Meng, Qi and Finley, Thomas and Wang, Taifeng and Chen, Wei and Ma, Weidong and Ye, Qiwei and Liu, Tie-Yan},
title = {LightGBM: a highly efficient gradient boosting decision tree},
year = {2017},
isbn = {9781510860964},
publisher = {Curran Associates Inc.},
address = {Red Hook, NY, USA},
booktitle = {Proceedings of the 31st International Conference on Neural Information Processing Systems},
pages = {3149–3157},
numpages = {9},
location = {Long Beach, California, USA},
series = {NIPS'17}
}

@article{petersen2011statistical,
  title={Statistical regularities in the rank-citation profile of scientists},
  author={Petersen, Alexander M. and Stanley, H. Eugene and Succi, Sauro},
  journal={Scientific Reports},
  volume={1},
  pages={181},
  year={2011},
  publisher={Nature Publishing Group},
  doi={10.1038/srep00181},
  url={https://doi.org/10.1038/srep00181}
}

@InProceedings{10.1007/978-3-319-07467-2_12,
author="Pobiedina, Nataliia
and Ichise, Ryutaro",
editor="Ali, Moonis
and Pan, Jeng-Shyang
and Chen, Shyi-Ming
and Horng, Mong-Fong",
title="Predicting Citation Counts for Academic Literature Using Graph Pattern Mining",
booktitle="Modern Advances in Applied Intelligence",
year="2014",
publisher="Springer International Publishing",
address="Cham",
pages="109--119",
}

@misc{mcinnes2020umapuniformmanifoldapproximation,
      title={UMAP: Uniform Manifold Approximation and Projection for Dimension Reduction}, 
      author={Leland McInnes and John Healy and James Melville},
      year={2020},
      eprint={1802.03426},
      archivePrefix={arXiv},
      primaryClass={stat.ML},
      url={https://arxiv.org/abs/1802.03426}, 
}

\clearpage
\pagebreak
\section*{Appendix A: prompts}
Here, we describe the prompts we passed to large language models to accomplish various tasks in the paper. To extract verifiable claims and investigation proposals from NSF grant abstracts we use the following system prompt:
\begin{lstlisting}
You are an expert at structured data extraction. You will be given unstructured text from a scientific abstract and should convert it into the given structure.
\end{lstlisting}

The full prompt to extract information from grant abstract is
\begin{lstlisting}
You are a scientific text analyzer specializing in NSF grant abstracts. Extract verifiable claims and investigation proposals from the given abstract.

DEFINITIONS:
- Verifiable claims: Statements presented as facts, established knowledge, or explicit/implicit assumptions that could be verified through existing evidence or prior research. Examples include:
  * Established scientific facts or findings
  * Current state of knowledge in the field
  * Existing problems or challenges
  * Prior research results referenced
  * Stated assumptions or premises

- Investigation proposals: Future research activities, hypotheses to test, questions to answer, or methods to develop that this award will pursue. Examples include:
  * Research questions to be addressed
  * Hypotheses to be tested
  * Methods to be developed or applied
  * Systems to be built or studied
  * Experiments to be conducted
  * Analyses to be performed

INSTRUCTIONS:
1. Read the entire abstract carefully
2. Identify ALL verifiable claims - there may be many throughout the abstract
3. Identify ALL investigation proposals - capture the complete research agenda
4. Keep each item concise but complete (typically 1-2 sentences)
5. Preserve technical terminology and specificity exactly as written
6. Extract only what is stated or directly implied - do not add interpretation
7. Maintain the distinction between current knowledge (claims) and future work (proposals)

EXAMPLE:
Input: Metal-organic frameworks (MOFs) have shown promise for CO2 capture due to their high surface areas and tunable pore structures. However, water stability remains a critical challenge for practical deployment. This project will develop new water-stable MOF architectures using hydrophobic ligands and investigate their CO2 adsorption performance under humid conditions. The team will also use machine learning to predict stability and guide synthesis.

Output:
{
  "verifiable_claims": [
    "Metal-organic frameworks (MOFs) have shown promise for CO2 capture due to their high surface areas and tunable pore structures",
    "Water stability remains a critical challenge for practical deployment of MOFs"
  ],
  "investigation_proposals": [
    "Develop new water-stable MOF architectures using hydrophobic ligands",
    "Investigate CO2 adsorption performance of new MOFs under humid conditions",
    "Use machine learning to predict MOF stability",
    "Use machine learning to guide MOF synthesis"
  ]
}

OUTPUT REQUIREMENTS:
- Include ALL claims and proposals found (there is no maximum limit)
- Use empty arrays [] if no claims or proposals are found

INPUT:
\end{lstlisting}

The full prompt to extract information from publications is
\begin{lstlisting}
You are a scientific text analyzer specializing in extracting key results from research publication abstracts. Extract verifiable scientific findings and reported methods from the given abstract.

DEFINITIONS:

- Scientific findings: Statements reporting discoveries, measurements, relationships, or conclusions based on the study. These are the results presented in the paper.
  Examples include:
  * Quantitative or qualitative outcomes
  * Newly identified patterns, mechanisms, or effects
  * Comparative or benchmarking results
  * Conclusions supported by evidence

- Reported methods: Specific techniques, procedures, models, data, or tools used to produce the findings.
  Examples include:
  * Experimental protocols
  * Modeling or simulation approaches
  * Data collection or analysis methods
  * Instruments or datasets used
  * Computational tools or algorithms applied

INSTRUCTIONS:

1. Read the entire abstract carefully  
2. Identify ALL scientific findings - what the paper reports as results  
3. Identify ALL reported methods - how those results were obtained  
4. Keep each item concise but complete (typically 1-2 sentences)  
5. Preserve technical terminology and specificity exactly as written  
6. Extract only what is explicitly stated or directly implied - do not speculate  
7. Clearly separate findings from methods

EXAMPLE:

Input:  
We demonstrate a new lithium-sulfur cathode design that improves cycling stability by 40% over conventional designs. Electrochemical impedance spectroscopy and scanning electron microscopy were used to characterize failure mechanisms. Our model shows that capacity fade correlates with polysulfide accumulation at the separator.

Output:
{
  "scientific_findings": [
    "A new lithium-sulfur cathode design improves cycling stability by 40% over conventional designs",
    "Capacity fade correlates with polysulfide accumulation at the separator"
  ],
  "reported_methods": [
    "Electrochemical impedance spectroscopy was used to characterize failure mechanisms",
    "Scanning electron microscopy was used to characterize failure mechanisms",
    "A model was used to correlate capacity fade with polysulfide accumulation"
  ]
}

OUTPUT REQUIREMENTS:
- Include ALL findings and methods found (there is no maximum limit)  
- Use empty arrays [] if no findings or methods are found

INPUT:
\end{lstlisting}

Finally, to calculate the ``success score'' between a specific proposal and a specific finding,
we use the following system prompt:
\begin{lstlisting}
You are an expert scientific evaluator. Your task is to score how well scientific findings achieve the research goals outlined in corresponding investigation proposals.
\end{lstlisting}

The main prompt is
\begin{lstlisting}
Main prompt: Rate 0-100 how well this scientific finding achieves the research goal in this proposal:
\end{lstlisting}

\end{document}